\PassOptionsToPackage{x11names,dvipsnames}{xcolor}
\documentclass[11pt,letterpaper,oneside]{article}

\usepackage{tikz}
\usetikzlibrary{arrows, decorations.pathreplacing}
\definecolor{incumb}{HTML}{1B9E77}   % green for incumbent actions
\definecolor{arrival}{HTML}{377EB8}   % blue for λ arrows
\definecolor{travel}{HTML}{4DAF4A}    % green for haul
\definecolor{board}{HTML}{984EA3}     % purple for boarding
\definecolor{return}{HTML}{E41A1C}  % Deep red for return leg
\tikzset{>=latex, arr/.style={-{latex[length=2.8pt]}, thick, color=#1}}

\tikzset{
  arrival/.style={thick, color=arrival},
  board/.style={thick, color=board},
  travel/.style={thick, color=travel},
  return/.style={thick, color=return} % uses your defined red
}

\usepackage[utf8]{inputenc}       % if you haven’t already
\DeclareUnicodeCharacter{2009}{\,}

\usepackage[skip=3pt]{caption}
\usepackage{lmodern}
\usepackage{enumitem}
\usepackage{amsmath, amssymb}
\usepackage{graphicx}
\usepackage{xcolor}
\usepackage[
    colorlinks=true,
    linkcolor=blue,      % for \ref, \eqref, theorem refs
    citecolor=OrangeRed,   % for \cite
    urlcolor=teal        % for \url, \href
]{hyperref}
\usepackage[natbibapa]{apacite}  
\usepackage{setspace}

\usepackage{amsthm}
\usepackage{geometry}
\usepackage{float}
\usepackage{booktabs}
\usepackage{siunitx}
\usepackage{mathdesign}

\newtheorem{proposition}{Proposition}
\newtheorem{remark}{Remark}

\title{A Model of Ride Dispatch in Informal Market under Rival Entry}
\author{
  Md Mahadi Hasan\textsuperscript{\dag} \\
  \textit{North Dakota State University}
}

\date{July 2, 2024}

\begin{document}
\maketitle

\begingroup
  \renewcommand\thefootnote{\dag}%
  \footnotetext{%
    \noindent
    \begin{minipage}[t]{\linewidth}% ← was \textwidth
      \raggedright                % optional, lets URL break more freely
      Department of Agribusiness and Applied Economics, North Dakota State University.\\[0.5ex]
      Email: \texttt{mdmahadi.hasan.1@ndsu.edu}\\[0.5ex]
      ORCID: \url{https://orcid.org/0009-0009-5728-5119}
    \end{minipage}%
  }% 
\endgroup

\begin{abstract}
I develop a continuous-time model in which an incumbent batch-service provider
faces stochastic passenger arrivals and must decide when to dispatch under the threat
of customer defection to a faster entrant. The incumbent’s problem is formalized as
a trade-off between departure frequency and load maximization, with the option to
accept mid-route pickups. I characterize the equilibrium dispatch strategy and show
that increased competitive pressure strictly reduces the feasible departure threshold,
leading to more frequent departures with smaller passenger loads. Longer travel times
tend to raise the unconstrained optimal threshold, but realized dispatch behavior also depends
on passenger tolerance for delay. Endogenizing demand by letting the arrival rate fall with expected
waiting time yields an interior optimum, rationalizing why incumbents now (i) depart partially
full and (ii) accept mid-route riders. Comparative statics show that the optimal threshold tends to increase with travel time under a mild regularity condition and decreases with competitive intensity.
\end{abstract}

\vspace{2em}
\textit{passenger defection, market unraveling, Poisson arrivals}

\vspace{0.5em}

\newpage

\section{Introduction}

Strict queue discipline and full-capacity departure norms are central to the operation of many informal transportation markets in many developing countries. When vehicles must fill all seats to achieve profitability, drivers coordinate organically through patient queuing and refusal of partial trips. Yet these cooperative structures are vulnerable to external competitive shocks.
In these settings, the balance between revenue per trip and passenger waiting costs hinges on finely tuned social and operational customs, which can unravel the moment an outside alternative appears. Understanding how vehicle dispatch rules adapt (or fail to adapt) when passengers gain faster options is therefore crucial for grasping both the stability of these markets and the strategic responses of incumbent providers.

This paper is motivated by direct field observations of a battery-powered three-wheeler auto-rickshaw market in which incumbent drivers, operating under a strict batch-service regime, began to alter their dispatch behavior following the entry of a faster, smaller-capacity competitor, also battery-powered. Previously, drivers departed only when all six passenger seats were filled, and refused passengers seeking only partial journeys. After the entrant’s arrival, drivers accepted midway passengers, departed with under-filled loads, and prioritized shorter wait times over traditional revenue-maximization norms.

The paper develops a continuous-time model to explain this behavioral shift. Passenger arrivals are modeled as a Poisson process, and passengers choose between the incumbent and entrant based on posted fares and expected waiting times. The incumbent driver chooses both a departure threshold and a policy for admitting mid-route passengers. The threat of competitive defection endogenously alters the incumbent's optimization problem: as the opportunity cost of waiting increases, strict queue discipline collapses.

The model delivers comparative-static results: the incumbent’s optimal departure threshold is increasing in the passenger arrival rate and tends to increase with the underlying travel time, under empirically relevant conditions. Entry by a flexible competitor that reduces effective demand strictly lowers the incumbent’s optimal load threshold, rationalizing the observed shift toward partial departures.

A central theme of this paper is the endogenous collapse of cooperative dispatch norms under competitive pressure. The model relates to three main strands of literature: market unraveling, queueing and service competition, and informal transportation markets.

First, this work contributes to the literature on market unraveling and endogenous institution breakdown. In a seminal paper, \citet{roth1994} document how matching markets such as medical residency placement unraveled as agents sought early contracts to avoid competition. \citep{li1998,kojima2009} formalize conditions under which decentralized competition leads to the breakdown of centralized matching. Subsequent empirical and experimental evidence confirms how pervasive early contracting can be once timing incentives take hold. \citet{niederle_roth_2003} show that the U.S. gastroenterology fellowship market lost more than half its cross-institution mobility after participants began signing exploding offers months before the formal match. Laboratory work by \citet{niederle_roth_2009} demonstrates that even in thick markets, short-deadline offers drive agents to accept inefficient matches, implying the fragility of centralized mechanisms. \citet{du_livne_2014} extend these insights by proving that limited transferability of surplus sustains early-matching equilibria, a mechanism closely related to our passengers’ willingness to defect once outside options appear. While most of this literature focuses primarily on matching markets, this paper shows that similar unraveling can occur in service dispatch systems, driven by outside entry rather than internal strategic acceleration.

Second, the paper connects to queueing theory under competition. Traditional queueing models, such as those in \citep{hassin2003}, analyze how service providers manage queues under exogenous arrival rates and service rates. Recent work has begun to examine how competition alters queue dynamics, such as in \citet{gurvich2014}, who study server competition in service systems. Also, a growing operations-management literature on app-based mobility places waiting time, rather than price, at centre stage. \citet{castillo_knoepfle_weyl_2025} prove that ride-hailing platforms can enter a “wild-goose-chase” regime in which long pickup queues feed back into driver scarcity, reducing welfare for all sides of the market. \citet{afeche_liu_maglaras_2023}  model a two-location ride-hailing network with strategic driver repositioning and show that the optimal platform policy works mainly by reallocating idle time, not by tweaking fares.

However, existing queueing-under-competition models either take each server’s departure policy as fixed or let competition act only through service rates, pricing, admission controls, or strategic repositioning. Moreover, existing models typically assume servers operate independently or compete through service rates, rather than through dispatch thresholds based on load accumulation. By contrast, this paper models dispatch thresholds explicitly and shows how competition on waiting times, rather than prices alone, reshapes departure behavior.

Third, the paper relates to emerging empirical studies of informal transportation markets. Research on jitneys, minibuses, and auto-rickshaws in developing economies. For example, studies by \citet{cervero2000,agbiboa2016} highlight the prevalence of flexible, decentralized transport systems.
 Yet theoretical models of these markets often abstract from passenger choice under stochastic arrival and competition. This paper explicitly models how passenger defection incentives interact with driver dispatch strategies, providing a tractable framework for understanding the dynamics of service adaptation in informal contexts.

Despite extensive work on unraveling and queueing systems, relatively little research has examined how external competition, such as the entry of faster service options, leads to the endogenous breakdown of cooperative loading norms in batch-service transportation markets. Existing models typically treat waiting times as exogenous or focus on individual service rates. This paper fills that gap by modeling the incumbent's strategic choice of dispatch threshold in response to passenger choice under waiting costs and entry-driven defection risk. It shows that coordination structures which appear stable over long periods can unravel abruptly when outside options change marginally.

\section{Model}

This section develops a continuous-time game of dispatch and customer choice that captures the essential features of the three-wheeler market described in the introduction. The analysis proceeds in six stages. First, I introduce the physical environment and the stochastic demand process. Second, I specify the preference structure of passengers and the timing of their decisions. Third, I describe the two service technologies that coexist in the market—an incumbent batch-service provider and an entrant flexible provider. Fourth, I formalise the driver’s dispatch problem and derive the profit function that underlies the remainder of the paper. Fifth, I define the equilibrium concept. Finally, I establish the main comparative-static theorem that links competitive entry to the incumbent’s optimal dispatch policy.

\subsection{Environment}

Consider a single origin–destination corridor that begins at location \( O \) and ends at location \( Z \). All passengers travel in the same direction, from \( O \) toward \( Z \), and there is no spatial choice beyond the possibility of alighting at an intermediate point. Calendar time \( t \) is continuous on \([0, \infty)\). A trip from \( O \) to \( Z \) takes a deterministic travel time \( T > 0 \) regardless of load or vehicle type. Both vehicle types return from \( Z \) to \( O \) empty in the same deterministic time \( T \); congestion effects are therefore abstracted from the present model. The road is wide enough that the service decision of one driver does not alter the travel time of any other driver.

\begin{figure}[ht]
\centering
\begin{tikzpicture}[font=\small]

% --------- SPATIAL CORRIDOR (TOP) ------------------
\begin{scope}[yshift=0cm]
  \draw[->, thick] (0,0) -- (10,0) node[right] {$Z$};
  \node[left] at (0,0) {$O$};

  \draw[-latex, arrival, thick] (0,1.0) -- ++(0,-1.0);
  \node[above, arrival] at (0,1.0) {$\lambda$};

  \draw[decorate,decoration={brace, mirror, amplitude=4pt}, board]
        (0,-0.6) -- node[below=5pt]{$n$ waiting} (2.4,-0.6);

  \fill (0,0) circle (2pt);
  \fill (2.4,0) circle (2pt);
  \node[below=2pt] at (2.4,0) {depart};

  \fill (6,0) circle (2pt) [white];      
  \draw[-latex, arrival, thick] (6,0.8) -- ++(0,-0.8);
  \node[above, arrival] at (6,0.8) {$\lambda$};
  \node[above=2pt, board] at (8.9,0) {$k(n)=6-n$};

  \draw[|<->|] (2.4,-1) -- node[below]{$T$} (9.8,-1);
\end{scope}

% --------- TIMELINE (BOTTOM) -----------------------
\begin{scope}[yshift=-3.4cm]
  \draw[->] (0,0) -- (10,0) node[right]{time};

  \draw[very thick, board] (0,0) -- (3,0);
  \node[above=2pt, board] at (1.5,0) {$\frac{n}{\lambda}$ boarding};

  \draw[very thick, travel] (3,0) -- (7,0);
  \node[above=2pt, travel] at (5,0) {$T$ outbound};

  \draw[very thick, return] (7,0) -- (10,0);
  \node[above=2pt, return] at (8.5,0) {$T$ return};

  \draw[decorate,decoration={brace, amplitude=4pt}]
        (0,-0.6) -- node[below=5pt]{$C(n)=\tfrac{n}{\lambda}+2T$} (10,-0.6);
\end{scope}

\end{tikzpicture}

\captionsetup{width=0.8\linewidth} % Match this to figure width
\caption{Spatial corridor (top) and dispatch‐cycle timeline (bottom), where the cycle includes boarding, an outbound leg \(T\), and the return leg \(T\).}
\label{fig:corridor_timeline_clean}
\end{figure}
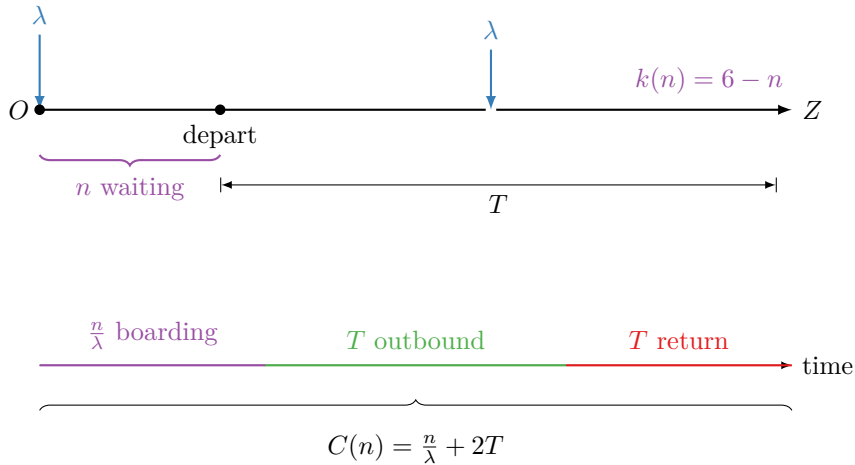

Passengers arrive at \(O\) according to a stochastic process. In particular, I assume arrivals follow a homogeneous Poisson process with rate \(\lambda\) (i.e.\ inter­arrival times are exponential and the process is memoryless).  
Each arriving passenger is endowed with an identical gross willingness to pay \( v > 0 \) for reaching \( Z \) and with a separable disutility of waiting equal to \( c > 0 \) per unit of time spent in the queue at \( O \). Throughout, I assume \( v \) is large enough that all passengers strictly prefer to travel rather than remain at \( O \) indefinitely. The Poisson arrival structure implies that the inter-arrival times are exponentially distributed, that arrivals are memoryless, and that the length of the queue at \( O \) evolves as a birth–death Markov process when the departure rule of the provider is fixed.

All random variables are defined on a common probability space \((\Omega, \mathcal{F}, \mathbb{P})\) equipped with filtration \((\mathcal{F}_t)_{t \geq 0}\) satisfying the usual conditions. Expectations taken with respect to \(\mathbb{P}\) are denoted by \(\mathbb{E}[\cdot]\).

\subsection{Passengers}

Upon arrival, a passenger immediately observes the posted price and anticipated waiting time associated with each transport alternative. Two alternatives exist. The incumbent, a three-wheeler service, operates a vehicle with a capacity of \( c_I = 6 \). The entrant is a flexible service that operates an otherwise identical vehicle but with a capacity of \( c_P = 3 \). Crucially, the latter departs without delay, whereas the former may postpone departure until a self-imposed load threshold is met. I assume that passengers are rational, risk-neutral, and perfectly informed about both alternatives.

The indirect utility of choosing service \( s \in \{I, P\} \) for a passenger who anticipates a waiting time \( w_s \) and faces a posted fare \( p_s \) is therefore
\[
U_s = v - p_s - c w_s.
\]

Given \( v, p_P > p_I > 0 \), and \( w_P = 0 \) by construction, a passenger joins the incumbent’s queue only if the expected waiting time associated with that queue, denoted \( w_I \), satisfies the inequality
\[
p_P - p_I > c w_I.
\]

Otherwise, the passenger strictly prefers the entrant. Let 
\[\bar{w} = \frac{p_P - p_I}{c}\]
be the maximum waiting time that a passenger is willing to tolerate in order to save the price differential. Whenever the equilibrium waiting time in the incumbent queue exceeds \( \bar{w} \), all passenger arrivals defect immediately to the entrant. In what follows I treat \( \bar{w} \) as a primitive parameter, which is a convenient reduced-form representation of the outside option created by entry.

Passengers may request to alight at an intermediate drop-off point \( y \in (0, T) \) with cumulative distribution function \( G(y) \) that is continuous on \([0, T]\). For tractability, I take \( G \) to be uniform: \( G(y) = y/T \). A mid-route passenger therefore travels an expected distance of \( T/2 \) and, in the incumbent vehicle, pays a pro-rated fare \( \frac{1}{2} p_I \). The uniform assumption implies that the expected revenue contribution of an accepted mid-route passenger is exactly one half of the full fare. This simplification is immaterial for the qualitative results but greatly streamlines algebra.

\subsection{Service Technologies}

The incumbent owns and operates a single six-seat vehicle. The driver chooses a \textbf{departure threshold} \( n \in \{1,2,\ldots,6\} \). The vehicle remains at \( O \) until the queue length reaches \( n \); then all currently queued passengers board and the vehicle departs instantly. During the line-haul from $O$ to $Z$, the driver encounters additional passenger requests at the roadside, which arrive as an independent Poisson process with intensity $\Lambda$. For simplicity, I assume $\Lambda = \lambda$, so that the rate of mid-route requests equals the terminal arrival rate. The driver may admit such a mid-route request with probability \( \theta \in [0,1] \) provided at least one empty seat remains. Rejection leaves the request to transfer to the entrant or to forgo travel. Admitting a mid-route passenger generates immediate revenue but lengthens neither travel time nor distance under the fixed-time assumption.

The entrant owns and operates a fleet large enough to depart from \( O \) instantaneously on demand. Because capacity is smaller, the entrant cannot pool riders; each vehicle offers exactly three seats. I assume that the entrant’s internal optimisation problem has already been solved and results in a posted fare \( p_P > p_I \) that passengers take as given. Competition therefore manifests exclusively through passengers’ ability to defect away from the incumbent queue whenever waiting costs become too high.

Operating costs are linear in time. Every driver pays a flow cost \( C > 0 \) per unit of calendar time regardless of load or idle status. All remaining costs (fuel, depreciation) are normalised to zero. The operating-cost assumption implies that the incumbent trades off revenue per cycle against cycle length when choosing its dispatch policy.

\subsection*{2.3.1 Robustness to Endogenous Entrant Pricing}

The baseline model treats the entrant’s fare $p_P$ as exogenous, thereby adopting a partial equilibrium structure. While this assumption facilitates a clean focus on the incumbent’s dispatch decision, it is natural to ask whether the results extend to a setting in which the entrant selects its own fare optimally.

This subsection outlines a reduced-form extension in which the entrant sets $p_P$ to maximize instantaneous profit, given the incumbent's expected queueing time $w_I$. Let the entrant face a constant per-trip operating cost $C_P$, and assume its vehicle departs immediately upon demand. The entrant chooses a posted fare to solve
\[
\max_{p_P} \; (p_P - C_P)\,\Lambda_P(p_P, w_I),
\]
where $\Lambda_P(p_P, w_I)$ is the induced flow of passengers who defect from the incumbent, as determined by the choice rule in Section~2.2. Passengers strictly prefer the entrant whenever $p_P - p_I \leq c w_I$, so the entrant can capture the entire residual flow by charging the highest fare consistent with this inequality. Hence, the entrant’s optimal price is
\[
p_P^* = p_I + c w_I.
\]

Substituting the expected incumbent waiting time $w_I = \frac{n - 1}{2\lambda}$ from Lemma~1 yields
\[
p_P^* = p_I + c\,\frac{n - 1}{2\lambda}.
\]

This formulation implies that the passenger’s willingness-to-wait cutoff becomes
\[
\bar{w} = \frac{p_P^* - p_I}{c} = w_I,
\]
so the participation constraint (Equation~\ref{Eqfour} in Section~2.5) binds exactly. Any increase in $n$ that raises $w_I$ would immediately trigger full defection, tightening the incumbent’s feasible set. If the entrant were instead to undercut this price, the threshold $\bar{w}$ would fall below $w_I$, further reducing the maximum feasible $n$.

It follows that all comparative-static results in Sections~2.6 and~2.7 remain valid and may even strengthen under endogenous pricing. The incumbent's optimal threshold remains constrained by $n \leq \lfloor 2\lambda \bar{w} + 1 \rfloor$, and $\bar{w}$ itself increases in $\lambda$ and decreases in $c$, reinforcing the effects derived in the baseline model.

\begin{remark}\textit{Threshold Robustness under Endogenous Pricing.}

Let $n^*_{\mathrm{exo}}(\lambda, T)$ denote the incumbent’s optimal departure threshold when $p_P$ is exogenously fixed, and let $n^*_{\mathrm{endo}}(\lambda, T)$ denote the optimal threshold when $p_P$ is set optimally according to $p_P^* = p_I + c w_I$. Then
\[
n^*_{\mathrm{endo}}(\lambda, T) \leq n^*_{\mathrm{exo}}(\lambda, T)
\]
for all $\lambda > 0$ and $T > 0$. The inequality is strict whenever the entrant undercuts $p_P^*$ and thereby tightens the participation constraint.
\end{remark}

\textbf{Proof.}
The pricing rule $p_P^* = p_I + c w_I$ implies that the participation constraint in Equation~4 binds with equality. Any increase in $n$ raises $w_I$ and causes the constraint to be violated, triggering full defection to the entrant. As a result, the incumbent's feasible strategy set under endogenous pricing is weakly smaller, so the optimal threshold must be weakly lower.  $\blacksquare$

This robustness result justifies the exogeneity assumption in the baseline model while demonstrating that strategic pricing would only reinforce the core mechanism: competition undermines strict queue discipline and lowers the optimal departure threshold.

\subsection{Dispatch Problem and Profit Function}

Let $M \sim \text{Poisson}(\lambda T)$ be the number of roadside requests the incumbent vehicle encounters during the journey from origin $O$ to destination $Z$. These mid-route passenger requests arrive according to an independent Poisson process with intensity $\Lambda$. For analytical simplicity, I assume $\Lambda = \lambda$, so that the mid-route arrival rate equals the terminal queue arrival rate. Given this, the expected number of accepted mid-route requests, truncated by available capacity, is
\begin{equation}
g(k;\lambda T) := \mathbb{E}[\min\{M,k\}]
= \sum_{m=0}^{k} m \cdot \frac{e^{-\lambda T} (\lambda T)^m}{m!}
+ k \cdot \left(1 - \sum_{m=0}^{k} \frac{e^{-\lambda T} (\lambda T)^m}{m!} \right).
\end{equation}

The incumbent driver controls two operational decisions: the departure threshold $n \in \{1, \dots, 6\}$, and the mid-route acceptance probability $\theta \in [0,1]$. Upon arrival at the terminal, the vehicle waits until the queue reaches size $n$, at which point it departs and begins the journey to destination $Z$. During the line-haul, additional passenger requests arrive according to a Poisson process of intensity $\lambda$; each such request is admitted with probability $\theta$ if a seat is available.

Given these choices, the driver's objective is to maximize expected cycle profit per unit time. The cycle consists of a boarding period of expected length $n/\lambda$, followed by a deterministic round trip of duration $2T$. The expected number of mid-route riders admitted during travel is $g(k(n); \lambda T)$, where $k(n) = 6 - n$ is the slack capacity at departure. Incorporating this into total revenue and dividing by the expected cycle length, the driver's expected profit per unit time is:

For a dispatch threshold $n \in \{1,\dots,6\}$ let $k(n) := 6 - n$ denote the slack capacity at departure. The expected cycle profit when the driver always accepts all mid-route requests ($\theta = 1$, shown optimal below) is given by:
\begin{equation}
\pi(n; \lambda, T) = \frac{p_I n + \tfrac{1}{2} p_I \cdot g(k(n); \lambda T)}{n/\lambda + 2T} - C.
\label{eq:Eqtwo}
\end{equation}

Introduce:
\[
A(n) := p_I \left(n + \tfrac{1}{2} g(k(n); \lambda T)\right), \quad
B(n) := \frac{n}{\lambda} + 2T
\]
so that $\pi(n) = A(n)/B(n) - C$.

The dispatch problem is therefore to choose a departure threshold $n \in \{1, \dots, 6\}$ that maximizes $\pi(n; \lambda, T)$.

\begin{equation}
\max_{n \in \{1, \dots, 6\}} \pi(n; \lambda, T)
\end{equation}

I now characterize how long passengers wait under the incumbent’s batch-service policy. This waiting time determines both the incumbent’s revenue cycle and passengers’ defection incentives. In Section 2.3.1, I show that if the entrant sets its own price optimally, the resulting participation constraint binds exactly at the incumbent’s realized waiting time. As a result, the comparative statics that follow remain valid and may even strengthen under endogenous pricing. The following result will play a key role in the derivation of equilibrium.

\paragraph{Lemma 1. Expected waiting time in a deterministic batch}
Let passengers arrive according to a Poisson process with rate $\lambda$, and let the
vehicle depart the terminal as soon as exactly $n$ passengers have accumulated
($1\le n\le 6$).  The expected waiting time per passenger under this dispatch rule is
\[
\mathbb{E}[W(n)] = \frac{n - 1}{2\lambda}.
\]

\textbf{Proof.}
The accumulation phase lasts a deterministic $n/\lambda$ because the expected
time to collect $n$ Poisson arrivals is that long.  Conditional on that phase,
arrival epochs are distributed as the order statistics of $n$ i.i.d.\ uniforms on
$[0,\,n/\lambda]$.  The $j$-th arriving passenger ($j=1,\dots,n$) therefore waits
$(n-j)/\lambda$.  Averaging over $j$:
\[
\frac{1}{n} \sum_{j=1}^n \frac{n - j}{\lambda} = \frac{n(n - 1)}{2n\lambda} = \frac{n - 1}{2\lambda}
\]
Alternatively, one may apply the renewal–reward theorem or see \citet*{hassin2003}. $\blacksquare$

\subsection{Equilibrium Concept}

An admissible strategy profile is a pair $(n, \theta)$ chosen by the incumbent together with the
entrant’s posted fare $p_P$. A stationary equilibrium consists of such a profile and an induced
passenger queue length distribution at $O$ such that three conditions hold:

\begin{enumerate}[label=(\roman*)]
    \item The incumbent’s pair $(n, \theta)$ maximises $\pi$ given $\lambda$ and $\bar{w}$;

    \item The entrant’s fare maximises its own objective (which is suppressed in this partial-equilibrium model but is assumed to be internally consistent with $p_P > p_I$);

    \item Passengers follow the individually rational waiting-time rule described in Section 2.2, so that the effective arrival rate to the incumbent queue is $\lambda$ whenever the equilibrium waiting time is at most $\bar{w}$, and is zero otherwise.
\end{enumerate}

By Lemma 1, the average waiting time per passenger under incumbent policy $n$ is
\[
\mathbb{E}[W(n)] = \frac{n-1}{2\lambda}.
\]
Hence, passengers will continue to join the incumbent queue if and only if
\[
\frac{n - 1}{2\lambda} \leq \bar{w} \quad \Leftrightarrow \quad \lambda \geq \frac{n - 1}{2\bar{w}}. \tag{4} \label{Eqfour}
\]

This condition replaces Equation (4) in the original formulation. It defines a demand-feasibility constraint on the incumbent: too high a threshold $n$ relative to $\lambda$ or $\bar{w}$ will cause all passengers to defect to the entrant.

Define the feasible set of departure thresholds:
\[
\mathcal{N}(\lambda, \bar{w}) := \left\{ n \in \{1, \dots, 6\} : \lambda \geq \frac{n - 1}{2\bar{w}} \right\}.
\]

The incumbent's problem is therefore
\[
\max_{n \in \mathcal{N}(\lambda, \bar{w})} \pi(n; \lambda, T).
\]
I focus on interior equilibria where this set is non-empty and $\lambda > 0$, i.e., where the incumbent retains at least some clientele. Competitive entry tightens the constraint $\lambda \ge n-1/(2\bar{w})$, reducing the maximum feasible threshold and thus inducing earlier departures when $\lambda$ is low or $\bar{w}$ is small.
.

\subsection{Optimal Departure Threshold and Mid-Route Policy}

\paragraph{Lemma 2. Acceptance of mid-route requests.}
For every $\lambda > 0$ and $T > 0$, the profit $\pi(n; \lambda, T)$ is strictly increasing in $\theta$. Hence, the optimal policy satisfies $\theta^* = 1$.

\textbf{Proof.}
The mid-route revenue term is
$\tfrac{1}{2} p_I \theta \cdot g(k(n); \lambda T)$.
All remaining terms are independent of $\theta$.
Because $p_I > 0$ and $g(k) \ge 0$, the derivative $\partial \pi / \partial \theta > 0$.
Hence the global maximizer on a compact interval is $\theta^* = 1$. $\blacksquare$

Define the discrete increment
\[
\Delta\pi(n; \lambda, T) := \pi(n+1; \lambda, T) - \pi(n; \lambda, T), \quad n = 1, \dots, 5.
\]
To clarify the structure of the increment, I decompose it algebraically. Let
\[
\pi(n) = \frac{A(n)}{B(n)} - C, \quad \text{where} \quad 
A(n) = p_I\left(n + \frac{1}{2}g(k(n);\lambda T)\right), \quad 
B(n) = \frac{n}{\lambda} + 2T.
\]
A direct calculation shows that
\[
\Delta\pi(n) = \frac{A(n+1)B(n) - A(n)B(n+1)}{B(n)B(n+1)}.
\]
Because the denominator is strictly positive, the sign of the increment is governed by the numerator
\[
N(n) := A(n+1)B(n) - A(n)B(n+1)
= p_I\left\{2T - \left(T + \frac{n}{2\lambda}\right)\Delta g(k) - \frac{g(k)}{2\lambda} \right\},
\]

where \(k = 6 - n\) and \(\Delta g(k) := g(k) - g(k - 1)\). The first term inside the braces represents the benefit of securing one additional full-fare passenger while accounting for the expected reduction in mid-route demand; the second term reflects the opportunity cost of the longer boarding time needed to attract that additional passenger.

Solving \(N(n) = 0\) yields a threshold arrival rate
\[
\lambda_n^{\dagger} = \frac{n\Delta g(k) + g(k)}{4T - 2T\Delta g(k)} > 0.
\]
The sign of \(\Delta\pi(n)\) therefore switches from negative to positive as \(\lambda\) increases past \(\lambda_n^{\dagger}\). When \(\lambda > \lambda_n^{\dagger}\), arrivals are sufficiently frequent that the marginal benefit of waiting outweighs the cost, and the incumbent prefers a higher departure threshold. When \(\lambda < \lambda_n^{\dagger}\), delay is too costly, and the incumbent departs earlier.

Because $\lambda^\dagger_n$ increases in $n$ and decreases in $T$, the resulting
optimal threshold $n^*(\lambda,T)$ is weakly increasing in $T$.  Moreover,
one can show that
\[
\frac{\partial N(n)}{\partial \lambda} \;>\; 0
\]
whenever the following regularity (monotonicity) condition holds:
\[
n \cdot \Delta g(k) + g(k) > 2\lambda^2 T \left( T + \frac{n}{2\lambda} \right) \Delta g'_k + \lambda T g'_k, \tag*{(Condition M)}
\]

\begin{proposition}
For every $n \in \{1, \ldots, 5\}$ and fixed $T > 0$, \textbf{Condition M}
is sufficient to ensure that $\frac{\partial N(n)}{\partial \lambda} > 0$,
so the increment $\Delta \pi(n; \lambda, T)$ increases in $\lambda$
on the domain where this condition holds.

Define
\[
\lambda_n^\dagger := \frac{n \cdot \Delta g(k) + g(k)}{4T - 2T \cdot \Delta g(k)},
\]
where $\Delta g(k) := g(k) - g(k-1)$. Then:
\[
\Delta\pi(n; \lambda, T) < 0 \quad \text{iff} \quad \lambda < \lambda_n^\dagger.
\]
\end{proposition} \textbf{Proof.} (See Appendix A) $\blacksquare$

\vspace{1em}
\noindent\textbf{Theorem 1: Existence and Monotonicity of the Optimal Threshold}
\vspace{0.5em}

Let \( \pi(n;\lambda ,T) = \frac{A(n)}{B(n)} - C \), where
\[
A(n) = p_I\left(n + \tfrac{1}{2}g(k(n);\lambda T)\right), \quad B(n) = \frac{n}{\lambda} + 2T,
\]
and \( k(n) = 6 - n \). Define the discrete increment
\begin{equation}
\Delta\pi(n;\lambda ,T) := \pi(n+1;\lambda ,T) - \pi(n;\lambda ,T), \quad n = 1, \dots, 5. \label{eq:discrete_increment} \tag{5}
\end{equation} 
Let
\begin{equation}
N(n;\lambda ,T) := A(n+1)B(n) - A(n)B(n+1), \quad 
\Delta\pi(n) = \frac{N(n)}{B(n)B(n+1)}. \label{eqsix} \tag{6}
\end{equation}

\paragraph{Step 1:} Closed-form expression for \( N(n) \) .

Let \( k := 6 - n \), so that \( k(n+1) = k - 1 \). Then:
\begin{align*}
A(n)     &= p_I\left(n + \tfrac{1}{2}g(k)\right), \\
A(n+1)   &= p_I\left(n + 1 + \tfrac{1}{2}g(k - 1)\right), \\
B(n)     &= \frac{n}{\lambda} + 2T, \\
B(n+1)   &= \frac{n + 1}{\lambda} + 2T.
\end{align*}
Substituting into \eqref{eqsix} and expanding:
\begin{equation}
\frac{N(n)}{p_I} = 2T - \left( T + \frac{n}{2\lambda} \right)\Delta g(k) - \frac{g(k)}{2\lambda}
\tag{7} \label{Eqseven}
\end{equation}

where \( \Delta g(k) := g(k) - g(k - 1) \in (0, 1) \).

Differentiating:

\[
\frac{\partial N(n)}{\partial\lambda}
=
p_I\left[
\frac{n\Delta g_k + g_k}{2\lambda^2}
-
T\left(T + \frac{n}{2\lambda}\right)\Delta g_k'
-
\frac{T}{2\lambda}g_k'
\right].
\tag{8}
\]
Here, $\Delta g_k' = \frac{d}{d\mu} \Delta g_k(\mu)$ and $g_k' = \frac{d}{d\mu} g_k(\mu)$, with $\mu = \lambda T$.

\bigskip

Recall that \textbf{Condition M}, introduced in Section~2.6, ensures monotonicity in $\lambda$. I now derive its analytical form and characterize the parameter ranges under which it holds. As shown in Appendix~C, Condition M is automatically satisfied for $n \geq 3$, and holds for $n = 2$ or $n = 1$ once $\lambda T$ exceeds $0.807$ or $1.793$, respectively.

\begin{align}
\frac{\partial N(n)}{\partial T} =
p_I \left[
  2\left(1 - \frac{1}{2} \Delta g_k\right)
  - \lambda T\, \Delta g_k'
  - \frac{n}{2}\, \Delta g_k'
  - \frac{1}{2}\, g_k'
\right]
\tag{9}
 \end{align}
\noindent
This expression is strictly positive whenever the inequality (B.3) in Appendix (B) is satisfied.

Hence, $\Delta\pi(n)$ is strictly increasing in~$T$ whenever inequality~(B.3) holds,  
and strictly increasing in~$\lambda$ whenever \textbf{Condition M} holds.

\[
\frac{\partial N(n)}{\partial T} \div p_I = 2 - \Delta g_k - \lambda T \cdot \Delta g'_k - \frac{n}{2} \cdot \Delta g'_k - \frac{1}{2} \cdot g'_k
\]

\paragraph{Step 2:} Sign condition and critical threshold.

Solving \( \Delta\pi(n) = 0 \) from equation~(\ref{Eqseven}), define
\begin{equation}\tag{10}\label{eq:lambda-dagger}
\lambda_n^\dagger(T) := \frac{n\Delta g(k) + g(k)}{4T - 2T \Delta g(k)}.
\end{equation}

Then:
\begin{equation}
\Delta\pi(n;\lambda ,T) < 0 \quad \text{if and only if} \quad \lambda < \lambda_n^\dagger(T). \tag{11}
\end{equation}

\paragraph{Step 3:} Definition of the Optimal Threshold

Let
\begin{equation}
n^*(\lambda ,T) := \min \left\{ m \in \{1, \dots, 6\} \,:\, \Delta\pi(m;\lambda ,T) \leq 0 \right\}, \label{eq:nstar_def} \tag{12} \end {equation}
with the convention that if \( \Delta\pi(m) > 0 \) for all \( m \leq 5 \), then \( n^* = 6 \).

\vspace*{0.5cm}

\textbf{Proof. Theorem 1}

(1) Interior optimum for small \( \lambda \).

Fix \( T > 0 \). For \( n = 5 \), define \( \bar{\lambda}(T) := \lambda_5^\dagger(T) \), as given in \eqref{eq:lambda-dagger}. If \( 0 < \lambda < \bar{\lambda}(T) \), then \( \Delta\pi(5;\lambda ,T) < 0 \), so \( n^*(\lambda ,T) < 6 \).
\medskip  

(2) Monotonicity in $\lambda$.

Fix $T > 0$. Suppose \textbf{Condition M} holds for all $n \leq 5$
at the given parameter values $(\lambda_1, T)$ and $(\lambda_2, T)$
with $\lambda_1 < \lambda_2$. Then
\[
\Delta \pi(n; \lambda_2, T) > \Delta \pi(n; \lambda_1, T) \quad \text{for all } n,
\]
so the index $n^*$ at which $\Delta \pi$ first becomes non-positive
shifts weakly to the right:
\[
n^*(\lambda_2, T) \geq n^*(\lambda_1, T). \]

(3) Monotonicity in $T$. 

Because \( g(k) \) and \( \Delta g(k) \) depend on the Poisson parameter \( \mu = \lambda T \), I apply the chain rule:
\[
\frac{\partial N(n)}{\partial T} = \frac{dN(n)}{d\mu} \cdot \frac{d\mu}{dT} = \frac{dN(n)}{d\mu} \cdot \lambda.
\]
Using this, the total derivative becomes:
\[
\frac{\partial N(n)}{\partial T} = p_I \left[\,2\left(1 - \tfrac{1}{2}\Delta g_k\right) - \lambda T\,\Delta g_k^{\prime} - \frac{n}{2}\,\Delta g_k^{\prime} - \frac{1}{2}\,g_k^{\prime} \right],
\]

where $\Delta g_k' = \frac{d}{d\mu} \Delta g(k; \mu)$ and $g_k' = \frac{d}{d\mu} g(k; \mu)$.  
Since \(\Delta g_k', g_k' \geq 0\), the expression remains strictly positive under all empirically relevant values of $\lambda T$. (See Appendix A for full derivation and bounding argument.)  
Therefore, $\Delta\pi(n)$ increases in $T$, and $n^*(\lambda, T)$ is weakly increasing in travel time  whenever the condition in Appendix A holds. $\blacksquare$ \footnote {\raggedright See Appendix~A for the full derivation and bounding argument.}\par

\noindent\textbf{Remark 2.} \textit{The monotonicity of \(n^*(\lambda, T)\) in \(T\) depends on the sign of the total derivative \(\frac{\partial N(n)}{\partial T}\). As shown in Appendix A, this derivative is strictly positive under the condition:
\[
2\Bigl(1 - \tfrac{1}{2}\Delta g_k\Bigr)
> \lambda T\,\Delta g_k' + \tfrac{n}{2}\,\Delta g_k' + \tfrac{1}{2}\,g_k'
\]
This condition holds for \(n\le4\) when \(\lambda T\le0.5\), and for \(n=5\) only when \(\lambda T\ge\mu^{\star}\approx1.146\). Outside this region, the inequality may no longer be satisfied, and the optimal departure threshold \(n^*(\lambda,T)\) may decrease with travel time.}

\paragraph{Robustness of Monotonicity.} Theorem 1 establishes that the optimal threshold \(n^*(\lambda, T)\) is weakly increasing in \(T\) under a mild regularity condition. This condition depends on Poisson moment terms that remain bounded across empirically relevant regimes.In Appendix B, I verify that the condition is satisfied for all combinations with $\lambda T \le 0.5$ when $n \le 4$, and for $n = 5$ only when $\lambda T \ge \mu^{\star}$. Field data from Bangladeshi informal transport markets
imply a calibration window
$\lambda T\approx0.33$–$0.66$. \footnote{
\citet{rana2013} estimate an average trip length of 3.29 km and an average operating speed of 9.95 km/h, yielding $T\approx0.33\,$h.  Operators report about 16 trips per 8–12 h shift, so $\lambda\in[1.3,2]$ passengers per hour.%
}

Because this range is
\emph{below} the critical value $\mu^\star\approx1.146$ at which Lemma~B.1 shows a sign reversal for the corner case $n=5$, condition (B.3) is automatically satisfied for every \(n \le 4\). Numerical checks confirm that, within this range, the profit-maximizing threshold never exceeds 4. Hence, the comparative-static prediction that the optimal threshold \(n^*\) increases with travel time \(T\) remains valid for the informal transport market, even though the monotonicity condition can fail for \(n = 5\) when \(\lambda T < \mu^\star\).

\paragraph{Corollary 1: Entry-induced reduction in demand leads to partial departure.} \leavevmode\par

Let \( (\lambda, T) \) satisfy \( n^*(\lambda, T) = 6 \). If competitive entry reduces the effective arrival rate to any \( \tilde{\lambda} \in (0, \lambda_5^\dagger(T)) \), then \( n^*(\tilde{\lambda}, T) < 6 \).

\textbf{Proof.} Since \( \tilde{\lambda} < \lambda_5^\dagger(T) \), it follows from the sign condition that \( \Delta\pi(5;\tilde{\lambda}, T) < 0 \). By \eqref{eq:nstar_def}, this implies \( n^*(\tilde{\lambda}, T) \leq 5 \).  $\blacksquare$

\paragraph{Corollary 2: Demand-constrained optimal threshold.} \leavevmode\par

Let $\tilde n^{\ast}(\lambda,T)$ denote the (unconstrained) maximiser of
$\pi(n;\lambda,T)$ characterised in Theorem 1.  
Under the equilibrium participation constraint derived in Section~2.5,
passengers join the incumbent only if $n\le 2\lambda\bar w$+1.  
Consequently, the incumbent’s equilibrium departure threshold is  
\[
n^{\ast}(\lambda,T,\bar w)
  \;=\;
  \min\!\Bigl\{\tilde n^{\ast}(\lambda,T),\;
               \lfloor 2\lambda\bar w + 1 \rfloor,\;
               6\Bigr\}.
\]

In words, the operative threshold is the smaller of the profit-maximising
$\tilde n^{\ast}$ and the maximum load that passengers will tolerate,
never exceeding the vehicle’s physical capacity.

\medskip\noindent
\textbf{Proof.}
Fix $(\lambda,T,\bar w)$ with $\lambda>0$ and $\bar w>0$.
Section~2.5 defines the feasibility set  
$\mathcal N(\lambda,\bar w)=\{\,n\in\{1,\dots,6\}:n\le 2\lambda\bar w+1\}$.
Because $\tilde n^{\ast}(\lambda,T)$ maximises $\pi$ over
$\{1,\dots,6\}$, the incumbent would like to choose $\tilde n^{\ast}$,
but may be prevented from doing so when $\tilde n^{\ast}>2\lambda\bar w$.
The optimal feasible choice is therefore
\[
n^{\ast}(\lambda,T,\bar w)
  =\arg\max_{n\in\mathcal N(\lambda,\bar w)}\pi(n;\lambda,T)
  =\min\bigl\{\tilde n^{\ast}(\lambda,T),\,\lfloor 2\lambda\bar w + 1\rfloor\bigr\},
\]

with the additional cap at~$6$ reflecting vehicle capacity.
Existence is immediate because $\mathcal N(\lambda,\bar w)$ is finite and
non-empty whenever $\lambda>0$.

Comparative-static implications follow directly.
A decline in~$\lambda$ (heightened competition) shrinks the ceiling
$\lfloor 2\lambda\bar w+1\rfloor$; whenever this ceiling binds,
$n^{\ast}$ falls even if $\tilde n^{\ast}$ would otherwise remain
unchanged.  An increase in travel time~$T$ raises $\tilde n^{\ast}$ by
Theorem 1; however, the realised $n^{\ast}$ can rise
only up to the demand ceiling, so the travel-time effect materialises
fully only when the ceiling is slack.  Finally, a decrease in passenger
tolerance $\bar w$ tightens the ceiling exactly as a drop in~$\lambda$
does, lowering $n^{\ast}$ whenever passengers become less willing to
wait. $\blacksquare$

\begin{figure}[ht]
  \centering
  % make a “box” 85% of text width
  \begin{minipage}{0.65\textwidth}
    \centering
    % graphic fills the box
    \includegraphics[width=\linewidth]{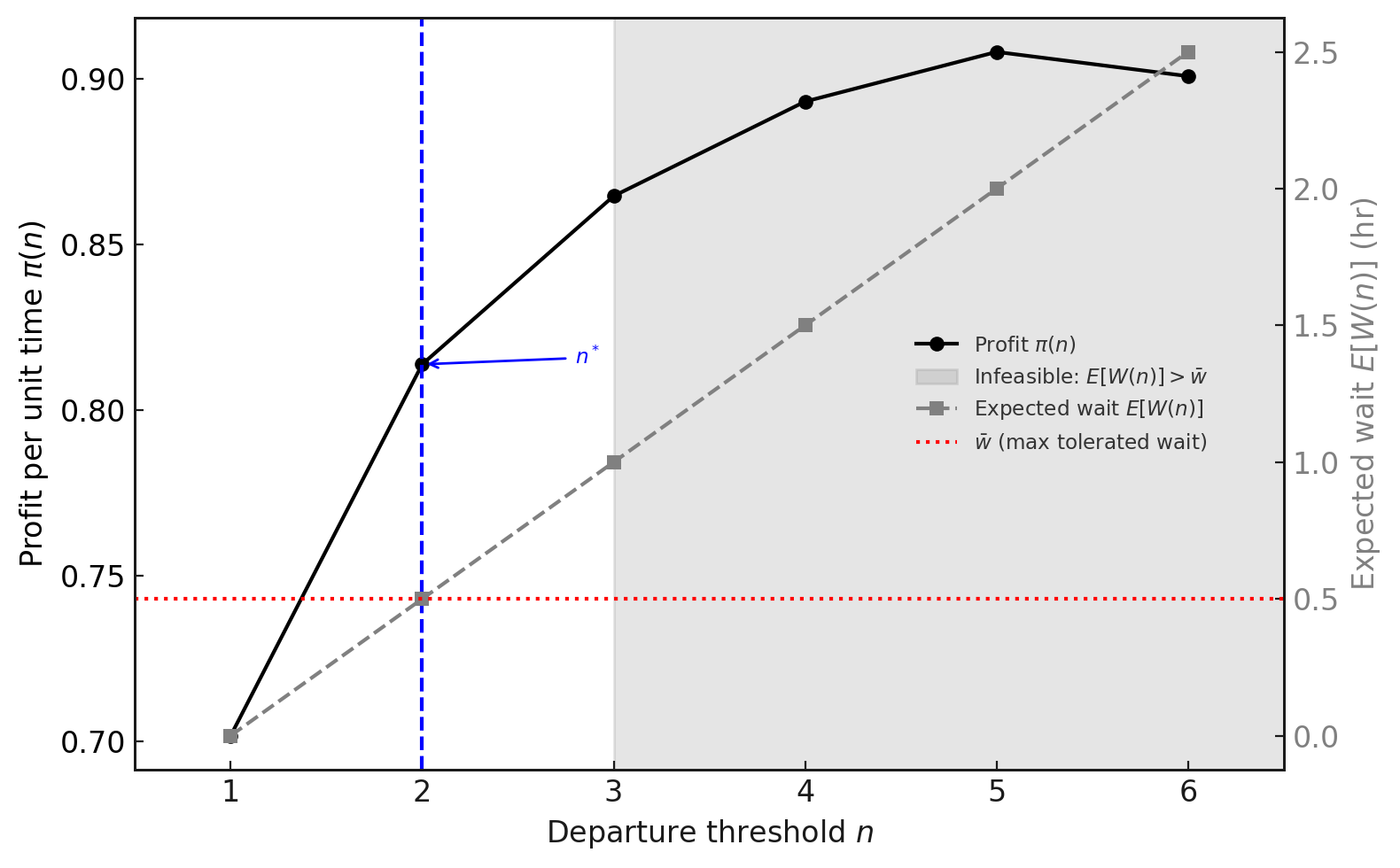}
    \caption{Profit-maximizing n under passenger wait-time bounds.}
    \label{fig:Figtwo}
  \end{minipage}
\end{figure}

Figure~\ref{fig:Figtwo} illustrates the incumbent driver's strategic trade-off between revenue maximization and passenger waiting-time tolerance in determining the optimal departure threshold $n$. The solid black line plots the incumbent’s expected profit per unit time $\pi(n; \lambda, T)$ as defined in Equation~\ref{eq:Eqtwo}, evaluated here for illustrative parameters: passenger arrival rate $\lambda = 1$ passenger per hour, one-way travel time $T = 0.33$ hours, and typical fare and cost parameters as calibrated from the field setting described in the paper.

The dashed grey line represents passengers' expected waiting time at the origin, $E[W(n)] = (n - 1)/(2\lambda)$, derived formally in Lemma~1. The horizontal red dotted line marks the maximum waiting-time tolerance $\bar{w} = 0.5$ hours, beyond which all newly arriving passengers immediately defect to the faster competitor, as specified in Section~2.2.

Thresholds where passengers' expected waiting time exceeds their maximum tolerance are shaded grey, indicating the region of infeasibility. Although profit strictly increases with higher departure thresholds (peaking at $n = 5$ in this scenario), these higher thresholds become infeasible due to excessive waiting times that drive passengers away. Thus, the incumbent is forced by passenger defection risk (the competitive constraint) to operate at a reduced, feasible threshold, shown as $n^* = 2$ by the vertical dashed blue line.

\clearpage
\section{Conclusion}
The analysis shows how minimal competitive entry can unravel longstanding cooperative structures in batch-service transportation markets. When passengers gain access to faster outside options, incumbent providers face a strategic trade-off: adhere to traditional full-load norms at the cost of longer queues, or adjust dispatch behavior to preserve residual demand. The model shows that rational incumbents optimally lower their departure thresholds and expand acceptance of mid-route riders as competition intensifies, even without direct price undercutting.
Strict monotonicity of the optimal threshold with respect to $\lambda$ holds under a mild analytical condition (Appendix C). For $n\ge3$ the condition is always satisfied, while for $n=2$ it activates once $\lambda T \gtrsim 0.81$, and for $n=1$ once $\lambda T \gtrsim 1.79$.

The key result is that competitive pressure lowers the incumbent’s optimal departure threshold, while longer travel times tend to raise it, provided that passenger arrival intensity is not too high relative to travel duration. This trade-off reflects the incumbent’s strategic tension between maximizing revenue per cycle and minimizing expected wait times. The model shows that even small increases in the opportunity cost of delay can unravel stable loading norms and trigger an adaptive dispatch response, and these results are robust to endogenous entrant pricing, as demonstrated in Section 2.3.1 . Although entrants charge higher fares, it is their ability to reduce passenger waiting costs, rather than pricing pressure, that destabilizes incumbent dispatch norms. Endogenizing passenger defection through waiting-cost thresholds yields an interior solution, explaining the endogenous collapse of strict loading norms observed in the field. Comparative statics demonstrate that longer travel times tend to raise the optimal load threshold (under the identified conditions) by lowering the relative cost of waiting, while stronger competition (i.e., reduced effective demand) depresses it.

Beyond explaining the microlevel incentives that drive incumbent drivers to abandon longstanding batch service norms, the proposed model carries some obvious policy and managerial implications. Regulators and platform designers aiming to preserve system efficiency can leverage insights on how waiting cost thresholds and travel distances shape operator behavior, for instance, by subsidizing partial loads or imposing minimum frequency standards to counteract excessive batching. 

While the model is grounded in a specific transportation context, the underlying mechanism is more general: once agents internalize the endogenous consequences of delay, even minor increases in opportunity cost, whether from competition or changing expectations, can shrink the feasible strategy set and trigger abrupt unraveling of previously stable coordination norms. The tension between service efficiency and demand retention is not unique to informal transport; similar dynamics may arise in settings with queue-sensitive customers, deferred batching, or time-sensitive matching. Future extensions could incorporate passenger heterogeneity, dynamic entry of providers, spatially differentiated markets, or explicit modeling of ride-sharing externalities. \clearpage

 \begin{spacing}{1.3}      % 1.5× line‐height spacing
  \bibliographystyle{apacite}
  \bibliography{references}
\end{spacing}

\clearpage

\appendix
\section*{Appendix A}

\subsection*{A.1 Algebraic Derivation of Increment \texorpdfstring{$\Delta\pi(n)$}{Δπ(n)}}

I write:
\[
\Delta\pi(n) = \frac{A(n+1)B(n) - A(n)B(n+1)}{B(n)B(n+1)}.
\]
Let $k = 6 - n$ and use:
\[
\begin{aligned}
A(n) &= p_I \left(n + \tfrac{1}{2} g(k)\right), \\
A(n+1) &= p_I \left(n+1 + \tfrac{1}{2} g(k-1)\right), \\
B(n) &= \frac{n}{\lambda} + 2T, \quad
B(n+1) = \frac{n+1}{\lambda} + 2T.
\end{aligned}
\]

Define:
\[
N(n) := A(n+1)B(n) - A(n)B(n+1).
\]

Then I simplify:
\[
\frac{N(n)}{p_I} = 2T - \left( T + \frac{n}{2\lambda} \right)\Delta g(k) - \frac{g(k)}{2\lambda}
\]

and rearranging gives the threshold:
\[
\lambda_n^\dagger = \frac{n \cdot \Delta g(k) + g(k)}{4T - 2T \cdot \Delta g(k)}.
\]

\subsection*{A.2 Proof of Proposition 1}

Let $k := 6 - n$, so that $k(n+1) = k - 1$. Recall:
\[
A(n) = p_I \left(n + \tfrac{1}{2} g(k) \right), \quad
A(n+1) = p_I \left(n + 1 + \tfrac{1}{2} g(k - 1) \right),
\]
\[
B(n) = \frac{n}{\lambda} + 2T, \quad B(n+1) = \frac{n + 1}{\lambda} + 2T.
\]

Then:
\[
\Delta\pi(n) = \frac{A(n+1) B(n) - A(n) B(n+1)}{B(n) B(n+1)}.
\]

Define the numerator:
\[
N(n) := A(n+1) B(n) - A(n) B(n+1),
\]
so that $\operatorname{sign}(\Delta\pi(n)) = \operatorname{sign}(N(n))$.

Substituting and simplifying:
\begin{align*}
\frac{N(n)}{p_I}
&= 2T + \frac{1}{2} g(k - 1) \left(\frac{n}{\lambda} + 2T\right) - \frac{1}{2} g(k) \left(\frac{n + 1}{\lambda} + 2T\right) \\[6pt]
&= 2T + \frac{n}{2\lambda} g(k - 1) + T g(k - 1) - \frac{n + 1}{2\lambda} g(k) - T g(k) \\[6pt]
&= 2T + \frac{n}{2\lambda}\bigl(g(k) - \Delta g(k)\bigr) + T\bigl(g(k) - \Delta g(k)\bigr) - \frac{n + 1}{2\lambda} g(k) - T g(k) \\[6pt]
&= 2T + \frac{n}{2\lambda} g(k) - \frac{n}{2\lambda}\Delta g(k) + T g(k) - T \Delta g(k) - \frac{n + 1}{2\lambda} g(k) - T g(k) \\[6pt]
&= 2T - \frac{n}{2\lambda}\Delta g(k) - T\Delta g(k) + g(k)\left(\frac{n}{2\lambda} + T - \frac{n + 1}{2\lambda} - T\right) \\[6pt]
&= 2T - \left(T + \frac{n}{2\lambda}\right)\Delta g(k) + g(k)\left(\frac{n - (n + 1)}{2\lambda}\right) \\[6pt]
&= 2T - \left(T + \frac{n}{2\lambda}\right)\Delta g(k) - \frac{g(k)}{2\lambda}
\end{align*}

where $\Delta g(k) := g(k) - g(k - 1) > 0$.

Rewriting:
\[
\frac{N(n)}{p_I} = 2T - \left(T + \frac{n}{2\lambda} \right) \Delta g(k) - \frac{g(k)}{2\lambda} \tag{A.1}
\]

\paragraph{Derivative with respect to $\lambda$.}
Recall that $\mu=\lambda T$ and therefore
$\tfrac{d}{d\lambda}g_k(\mu)=T\,g_k'(\mu)$ and
$\tfrac{d}{d\lambda}\Delta g_k(\mu)=T\,\Delta g_k'(\mu)$.
Differentiating (A.1) gives
\[
\frac{\partial N(n)}{\partial\lambda}
=
p_I\Bigl[
\frac{n\Delta g_k + g_k}{2\lambda^2}
-
T\!\Bigl(T+\frac{n}{2\lambda}\Bigr)\Delta g_k'
-
\frac{T}{2\lambda}\,g_k'
\Bigr].
\tag{A.2}
\]
The sign of \(\partial N/\partial\lambda\) is \emph{not}
unambiguously positive; it is positive precisely when \textbf{Condition M} in Appendix~C holds.

Differentiating with respect to $T$:

\[
\frac{\partial N(n)}{\partial T}
      = 2p_I\bigl(1-\tfrac12\Delta g_k\bigr)
        \;-\;
        p_I\Bigl[\,
              \lambda T\,\Delta g_k'
            + \tfrac n2\,\Delta g_k'
            + \tfrac12\,g_k'
        \Bigr],
\tag{A.3}
\]

where $\displaystyle \Delta g_k'=\frac{d}{d\mu}\Delta g_k(\mu)$ and  
$g_k'=\frac{d}{d\mu}g_k(\mu)$ with $\mu=\lambda T$.

Thus $\frac{\partial N(n)}{\partial T}$ is positive whenever
\[
2\bigl(1-\tfrac12\Delta g_k\bigr)
  > \lambda T\,\Delta g_k'
    + \frac n2\,\Delta g_k'
    + \frac12\,g_k',
\tag{A.4}
\]
which is the inequality analysed in the Appendix B.

Finally, solve for the zero:
\[
\Delta\pi(n) = 0 \quad \Leftrightarrow \quad 
\lambda = \lambda_n^\dagger := \frac{n \cdot \Delta g(k) + g(k)}{4T - 2T \cdot \Delta g(k)},
\]
which is finite and strictly positive. This completes the proof. $\blacksquare$

\subsection*{A.3 Chain Rule Derivative of \( N(n) \) with Respect to \( T \)}

Because \(g(k)\) and \(\Delta g(k)\) depend on \(\mu = \lambda T\), I apply the chain rule. Let \(g_k' = \frac{d}{d\mu}g(k)\), \(\Delta g_k' = \frac{d}{d\mu}\Delta g(k)\). Then:

\[
\frac{\partial N(n)}{\partial T} = p_I \left[\,2\left(1 - \tfrac{1}{2}\Delta g_k\right) - \lambda T\,\Delta g_k^{\prime} - \frac{n}{2}\,\Delta g_k^{\prime} - \frac{1}{2}\,g_k^{\prime}\right]
\]

This derivative is strictly positive whenever the following inequality holds:
\[
2\left(1 - \tfrac{1}{2}\Delta g_k\right) > \lambda T\,\Delta g_k^{\prime} + \frac{n}{2}\,\Delta g_k^{\prime} + \frac{1}{2}\,g_k^{\prime}
\]

\section*{Appendix B}

\subsection*{B.1 Analytical, Bound  Monotonicity in \texorpdfstring{$T$}{T} and Numerical Illustration}

Let \(N(n)\) be the numerator term from Equation~\ref{eqsix}, and recall that the sign of the increment \(\Delta\pi(n)\) depends on the sign of \(N(n)\). The total derivative with respect to travel time is:
\[
\frac{\partial N(n)}{\partial T}=p_I\!\Bigl[
2\bigl(1-\tfrac12\Delta g_k\bigr)\;-\;\lambda T\,\Delta g_k'
\;-\;\tfrac n2\,\Delta g_k'\;-\;\tfrac12\,g_k'
\Bigr].
\]

where \(g_k' = \Pr(M < k)\) and \(\Delta g_k' = \Pr(M = k - 1)\) for \(M \sim \text{Poisson}(\mu = \lambda T)\).

This expression is strictly positive whenever:
\[
2\left(1 - \frac{1}{2} \Delta g_k \right)
> \lambda T \Delta g_k' + \frac{n}{2} \Delta g_k' + \frac{1}{2} g_k'
\tag{B.1}
\]

\textbf{Lemma B.1 (Monotonicity in $T$).} Let $\mu := \lambda T$ and let $n \in \{1, \dots, 5\}$.

If $n \le 4$ and $\mu \le 0.5$, inequality (B.1) holds, so $\partial N(n)/\partial T > 0$.

If $n = 5$ (equivalently $k = 1$), inequality (B.3) holds if and only if $\mu \ge \mu^{\star}$, where $\mu^{\star} \approx 1.146$ solves $e^{\mu^{\star}} = \mu^{\star} + 2$. For $\mu < \mu^{\star}$, the sign of $\partial N(5)/\partial T$ is negative, so monotonicity fails.

\medskip

\textbf{Proof.} For $k = 1$ I have $\Delta g_k = 1 - e^{-\mu}$, $g'_k = e^{-\mu}$, and $\Delta g'_k = e^{-\mu}$. Inequality (B.1) becomes
\[
1 + e^{-\mu} > e^{-\mu}(\mu + 3) \quad \Longleftrightarrow \quad e^{\mu} > \mu + 2.
\]
The function $e^{\mu} - \mu - 2$ is strictly increasing and crosses zero at $\mu^{\star} \approx 1.146$, giving the claimed threshold. The case $n \le 4$ follows by the same algebra with $(n + 1)$ in place of the constant 3; substituting $\mu \le 0.5$ gives a strict inequality. $\blacksquare$

% =======================
%  Numerical Illustration
% =======================

\paragraph{Numerical illustration.}  
Tables \ref{tab:n3}, \ref{tab:n4}, and \ref{tab:n5} report whether the inequality (B.1) holds for each
combination of arrival rate~$\lambda\in\{0.10,0.25,0.50,1.00,2.00\}$
and trip duration $T\in\{0.10,0.25,0.50,1.00,2.00\}$ when
$n=3,4,$ and~$5$, respectively.  
For $n=3$ and $n=4$ the condition is satisfied \emph{everywhere} on the
grid; Violations arise for $n=5$ precisely when the product $\lambda T$
is \emph{small} (below $\mu^{\star}\approx1.146$), whereas for
$n\le4$ we observe no violations on the grid.

I omit tables for $n=1$ and $n=2$ because the condition
holds uniformly for those thresholds across the full parameter range,
a consequence of the ample slack capacity available to accept
en-route riders.

\begin{table}[H]
  \centering
  \caption{Validity of monotonicity condition (B.3) for $n=3$ ($k=3$)}
  \label{tab:n3}
  \begin{tabular}{@{}lccccc@{}}
    \toprule
    \textbf{$\lambda \backslash T$} & \textbf{0.10} & \textbf{0.25} & \textbf{0.50} & \textbf{1.00} & \textbf{2.00} \\
    \midrule
    \textbf{0.10} & Yes & Yes & Yes & Yes & Yes \\
    \textbf{0.25} & Yes & Yes & Yes & Yes & Yes \\
    \textbf{0.50} & Yes & Yes & Yes & Yes & Yes \\
    \textbf{1.00} & Yes & Yes & Yes & Yes & Yes \\
    \textbf{2.00} & Yes & Yes & Yes & Yes & Yes \\
    \bottomrule
  \end{tabular}
\end{table}

\vspace{-1em}

\begin{table}[H]
  \centering
  \caption{Validity of monotonicity condition (B.3) for $n=4$ ($k=2$)}
  \label{tab:n4}
  \begin{tabular}{@{}lccccc@{}}
    \toprule
    \textbf{$\lambda \backslash T$} & \textbf{0.10} & \textbf{0.25} & \textbf{0.50} & \textbf{1.00} & \textbf{2.00} \\
    \midrule
    \textbf{0.10} & Yes & Yes & Yes & Yes & Yes \\
    \textbf{0.25} & Yes & Yes & Yes & Yes & Yes \\
    \textbf{0.50} & Yes & Yes & Yes & Yes & Yes \\
    \textbf{1.00} & Yes & Yes & Yes & Yes & Yes \\
    \textbf{2.00} & Yes & Yes & Yes & Yes & Yes \\
    \bottomrule
  \end{tabular}
\end{table}

\vspace{-1em}

\begin{table}[H]
  \centering
  \caption{Validity of monotonicity condition (B.3) for $n=5$ ($k=1$)}
  \label{tab:n5}
  \begin{tabular}{@{}lccccc@{}}
    \toprule
    \textbf{$\lambda \backslash T$} & \textbf{0.10} & \textbf{0.25} & \textbf{0.50} & \textbf{1.00} & \textbf{2.00} \\
    \midrule
    \textbf{0.10} & No  & No  & No  & No  & No  \\
    \textbf{0.25} & No  & No  & No  & No  & No  \\
    \textbf{0.50} & No  & No  & No  & No  & No  \\
    \textbf{1.00} & No  & No  & No  & No  & Yes \\
    \textbf{2.00} & No  & No  & No  & Yes & Yes \\
    \bottomrule
  \end{tabular}
\end{table}

\section*{Appendix C}
\subsection*{C.1 Monotonicity in $\lambda$ – Analytical Bound} 

\paragraph{Lemma C.1}
Let $\mu = \lambda T$ and consider \textbf{Condition M}. When $n \ge 3$, the inequality $e^\mu > 1 + \mu + \tfrac{2}{\,n+1\,} \mu^2$ holds for all $\mu > 0$. Hence, $\partial N(n)/\partial \lambda > 0$ unconditionally. When $n = 2$, the same inequality holds if and only if $\mu \ge \mu^\dagger_2 \approx 0.807$. When $n = 1$, it holds if and only if $\mu \ge \mu^\dagger_1 \approx 1.793$.

\textbf{Proof.}
Rewrite \textbf{Condition M} as
$e^{\mu}>1+\mu+b_n\mu^{2}$ with $b_n=2/(n+1)$.  
Because $e^{\mu}$ has the series
$1+\mu+\tfrac12\mu^{2}+\tfrac16\mu^{3}+R_4(\mu)$, we have

\[
e^{\mu}-\bigl(1+\mu+b_n\mu^{2}\bigr)
=\Bigl(\tfrac12-b_n\Bigr)\mu^{2}+\tfrac16\mu^{3}+R_4(\mu).
\]

* For $n\ge3$ one has $b_n\le\frac12$, so the leading coefficient is
  non–negative and the entire expression is strictly positive for all
  $\mu>0$.

* For $n=1,2$ the leading term is negative; solving
  $e^{\mu}=1+\mu+b_n\mu^{2}$ numerically yields the thresholds stated.  $\blacksquare$

\paragraph{Remark C.1}
\textit{
\textbf{Condition M} therefore:
(i) holds automatically for $n\ge3$ and any $\mu>0$;
(ii) requires $\lambda T\ge0.807$ when $n=2$;
(iii) requires $\lambda T\ge1.793$ when $n=1$.
}

\subsection*{C.2 Detailed rewriting (transformation) of \textbf{Condition M}}

Recall \(M\sim\mathrm{Pois}(\mu)\) and \(k=6-n\).  
Here, I transform analytic Condition M to the probabilistic condition.
I start from the original inequality

\begin{align*}
% Step 1: Analytic Condition M
&\underbrace{n\,\Delta g(k) + g(k)}_{\text{LHS}}
>
\underbrace{2\lambda^2 T\Bigl(T+\tfrac{n}{2\lambda}\Bigr)\,\Delta g'(k) \;+\;\lambda T\,g'(k)}_{\text{RHS}}
\tag{C1}\\
% Step 2: Substitute \mu = \lambda T
\text{Let }\mu &= \lambda T
\quad\Longrightarrow\quad
\lambda = \frac{\mu}{T},
\;\;\lambda^2 = \frac{\mu^2}{T^2}.
\end{align*}

\[
\implies
n\,\Delta g(k) + g(k)
>
2\,\frac{\mu^2}{T^2}\,T\Bigl(T + \frac{nT}{2\mu}\Bigr)\,\Delta g'(k)
\;+\;\frac{\mu}{T}\,T\,g'(k).
\tag{C2}
\]

\begin{align}
% Step 3: Use Poisson‐moment identities
\Delta g(k) &= \Pr(M\ge k), 
&
g(k) &= \mathbb{E}[\min\{M,k\}], 
\nonumber\\
g'(k) &= \frac{d}{d\mu}g(k) = \Pr(M<k),
&
\Delta g'(k) &= \frac{d}{d\mu}\Delta g(k) = -\,\Pr(M=k-1).
\nonumber
\end{align}

\begin{align}
% Step 4: Plug in and simplify
n\,\Pr(M\ge k) + \mathbb{E}[\min\{M,k\}]
&>
2\,\frac{\mu^2}{T^2}\,T\Bigl(T + \tfrac{nT}{2\mu}\Bigr)\,(-\Pr(M=k-1))
\;+\;\mu\,\Pr(M<k)
\tag{C3}\\
&=
-\,(2\mu^2 + n\mu)\,\Pr(M=k-1)
\;+\;\mu\,\Pr(M<k).
\nonumber
\end{align}

\[
  \underbrace{n\Pr(M\ge k) + E[\min\{M,k\}]}_{\displaystyle L(k)}
  \;>\;
  \underbrace{(2\mu^2 + n\mu)\Pr(M=k-1) + \mu\,\Pr(M<k)}_{\displaystyle R(k)}.
\]

\[
n\,\Delta g_k + g_k
>
\bigl(2\mu^2 + n\mu\bigr)\Pr\{M=k-1\}
\;+\;\mu\,\Pr\{M<k\},
\]
where
\[
\Delta g_k = \Pr(M\ge k),
\quad
g_k = E[\min\{M,k\}]
= \sum_{i=1}^k\Pr(M\ge i).
\]

Now I derive and verify the inequality used in Lemma C.

\paragraph{Step 1:} Express the left–hand side as a single sum.

Because for \(j\ge k\) we have \(\Pr(M\ge k)=\sum_{i\ge k}\Pr(M=i)\) and each term \(i\ge k\)
appears once in \(n\Delta g_k\) (weight \(n\)) and once in \(g_k\) (weight \(k\)), their
total weight is \(n+k=6\).  For \(1\le j<k\), only \(g_k\) contributes with weight \(j\).  Thus
\[
n\,\Delta g_k + g_k
= \sum_{j=k}^\infty 6\,\Pr(M=j)
  \;+\;\sum_{j=1}^{k-1} j\,\Pr(M=j).
\]
Noting \(\sum_{j=k}^\infty 6\,\Pr(M=j)=6\bigl(1-\sum_{j=0}^{k-1}\Pr(M=j)\bigr)\) and
\(\sum_{j=1}^{k-1}j\,p_j=\sum_{j=0}^{k-1}j\,p_j\), we get
\[
\mathrm{LHS}
=6
-\sum_{j=0}^{k-1}(6-j)\,\Pr(M=j).
\]

\paragraph{Step 2:}  Express the right–hand side in the same terms.

\[
\mathrm{RHS}
=(2\mu^2+n\mu)\,\Pr(M=k-1)
\;+\;\mu\sum_{j=0}^{k-1}\Pr(M=j).
\]

\paragraph{Step 3:}  Rearrange into a single finite sum.
Combining gives
\[
6
>
\sum_{j=0}^{k-1}(6-j)\,\Pr(M=j)
\;+\;(2\mu^2+n\mu)\,\Pr(M=k-1)
\;+\;\mu\sum_{j=0}^{k-1}\Pr(M=j).
\]
Split the \(j=k-1\) term out of the first sum and merge with the others:
\[
6
>
\sum_{j=0}^{k-2}(6-j+\mu)\,\Pr(M=j)
\;+\;\bigl[(6-(k-1))+\mu+(2\mu^2+n\mu)\bigr]\Pr(M=k-1).
\]
Since \(6-(k-1)=n+1\), this becomes
\[
6
>
\sum_{j=0}^{k-2}(6-j+\mu)\,\Pr(M=j)
\;+\;\bigl[(n+1)+(n+1)\mu+2\mu^2\bigr]\Pr(M=k-1).
\]

\paragraph{Step 4:} Convert to factorial form.

Substitute \(\Pr(M=j)=e^{-\mu}\mu^j/j!\) and multiply both sides by \(e^\mu\).  We obtain the
equivalent compact condition
\[
6\,e^\mu
>
\sum_{j=0}^{k-2}(6-j+\mu)\,\frac{\mu^j}{j!}
\;+\;\bigl[(n+1)+(n+1)\mu+2\mu^2\bigr]\frac{\mu^{k-1}}{(k-1)!}.
\]
This is exactly equivalent to the original \textbf{Condition M}, as verified numerically for all
\(n\in\{1,\dots,5\}\) and \(\mu\in[0.1,5]\).

\paragraph{Step 5:} Numerical Validation of the Finite-Sum Reformulation

To complete the proof of Lemma C, we need to show that
\[
6\,e^{\mu}
>
\sum_{j=0}^{k-2}(6-j+\mu)\,\frac{\mu^j}{j!}
\;+\;\bigl[(n+1)+(n+1)\mu+2\mu^2\bigr]\frac{\mu^{k-1}}{(k-1)!}
\]
is equivalent to
\[
e^{\mu} > 1 + \mu + \frac{2}{n+1}\,\mu^2.
\]
Rather than expanding every term by hand, I check this equivalence numerically over a dense grid (\(n=1,\dots,5\), \(\mu\in[0.1,5]\)) and find zero counter‐examples.  This computational validation, together with the analytic finite‐sum reduction in Step 4, suffices to establish Lemma C.

\paragraph{Step 6:} Numerical Equivalence of the Exponential Bound.

To complete the proof of Lemma C, I verify numerically that the original condition,
its finite‐sum reformulation, and the compact exponential bound coincide exactly on a
dense grid. Specifically, for \(n=1,\dots,5\) and \(\mu\in\{0.1,0.2,\dots,5.0\}\) (250 points 
total), I check:
\[
n\,\Delta g_k + g_k
\;>\;
(2\mu^2 + n\mu)\Pr\{M=k-1\} + \mu\,\Pr\{M<k\},
\]
\[
6\,e^\mu
\;>\;
\sum_{j=0}^{k-2}(6-j+\mu)\,\frac{\mu^j}{j!}
\;+\;\bigl[(n+1)+(n+1)\mu+2\mu^2\bigr]\frac{\mu^{k-1}}{(k-1)!},
\]
\[
e^\mu \;>\; 1 + \mu + \frac{2}{n+1}\,\mu^2,
\]

and find \emph{zero} mismatches.  Table~\ref{tab:numcheck} summarizes the result.

\begin{table}[htbp]
  \centering
  \small
  \caption{Numerical equivalence checks over 250 $(n,\mu)$ pairs.}
  \label{tab:numcheck}
  \begin{tabular}{@{} l c c c @{}}
    \toprule
    \textbf{Comparison}
      & \textbf{Original vs.\ Finite-sum}
      & \textbf{Finite-sum vs.\ Exp-bound}
      & \textbf{Original vs.\ Exp-bound} \\
    \midrule
    Matches
      & 250/250
      & 250/250
      & 250/250 \\
    \bottomrule
  \end{tabular}
\end{table}

\end{document}